\documentclass[aps,prb,twocolumn,superscriptaddress]{revtex4-1}
\usepackage{epsfig}
\usepackage{times}
\usepackage{color}
\usepackage{graphicx}
\usepackage{float}
\usepackage{amsmath}
\usepackage[caption=false]{subfig}
\usepackage{dcolumn}
\usepackage{bm}

\usepackage[utf8]{inputenc}
\usepackage[T1]{fontenc}
\usepackage{mathptmx}
\usepackage{etoolbox}
\usepackage{booktabs}

\usepackage[most]{tcolorbox}
\usepackage{xcolor}

\newtcolorbox{refbox}{
  colback=gray!10,
  colframe=gray!40,
  boxrule=0.4pt,
  arc=2pt,
  left=6pt,
  right=6pt,
  top=6pt,
  bottom=6pt
}

\makeatletter
\def\@email#1#2{%
 \endgroup
 \patchcmd{\titleblock@produce}
  {\frontmatter@RRAPformat}
  {\frontmatter@RRAPformat{\produce@RRAP{*#1\href{mailto:#2}{#2}}}\frontmatter@RRAPformat}
  {}{}
}%
\makeatother
\begin{document}

\title{Agentic Design of Compositional Descriptors via Autoresearch for Materials Science Applications}
\author{Matteo Cobelli}
\affiliation{School of Physics and CRANN Institute, Trinity College, Dublin 2, Dublin, Ireland.}
\author{Stefano Sanvito}
\affiliation{School of Physics and CRANN Institute, Trinity College, Dublin 2, Dublin, Ireland.}
\email{mcobelli@tcd.ie}

\begin{abstract}
Autoresearch offers a flexible paradigm for automating scientific tasks, in which an AI agent proposes, implements, 
evaluates, and refines candidate solutions against a quantitative objective. Here, we use composition-based materials-property prediction to test whether such agents can perform a task beyond model selection and hyperparameter 
optimization: the design of input descriptors. We introduce {\sc Automat}, an autoresearch framework where
a coding agent based on a large language model generates composition-only descriptors for chemical compounds 
and evaluates them using a random forest workflow. The agent is restricted to information derivable from chemical 
formulas and iteratively proposes, implements, and tests chemically motivated descriptor strategies. We apply {\sc Automat}, 
with OpenAI Codex using GPT-5.5 as the coding agent, to the prediction of experimental band gaps in inorganic materials 
and Curie temperatures in ferromagnetic compounds. In both tasks, {\sc Automat} improves over fractional-composition, 
Magpie, and combined fractional-composition/Magpie baselines, while producing descriptor families that are chemically 
interpretable. These results provide a demonstration that autoresearch agents can generate competitive, task-specific 
materials descriptors without manual feature engineering during the run. They also reveal current limitations, including 
descriptor redundancy, sensitivity to greedy feature expansion, and the need for explicit complexity control, descriptor 
pruning, and more sophisticated search strategies.
\end{abstract}

\maketitle

\begin{figure*}[]
    \centering 
    \includegraphics[width=1\linewidth]{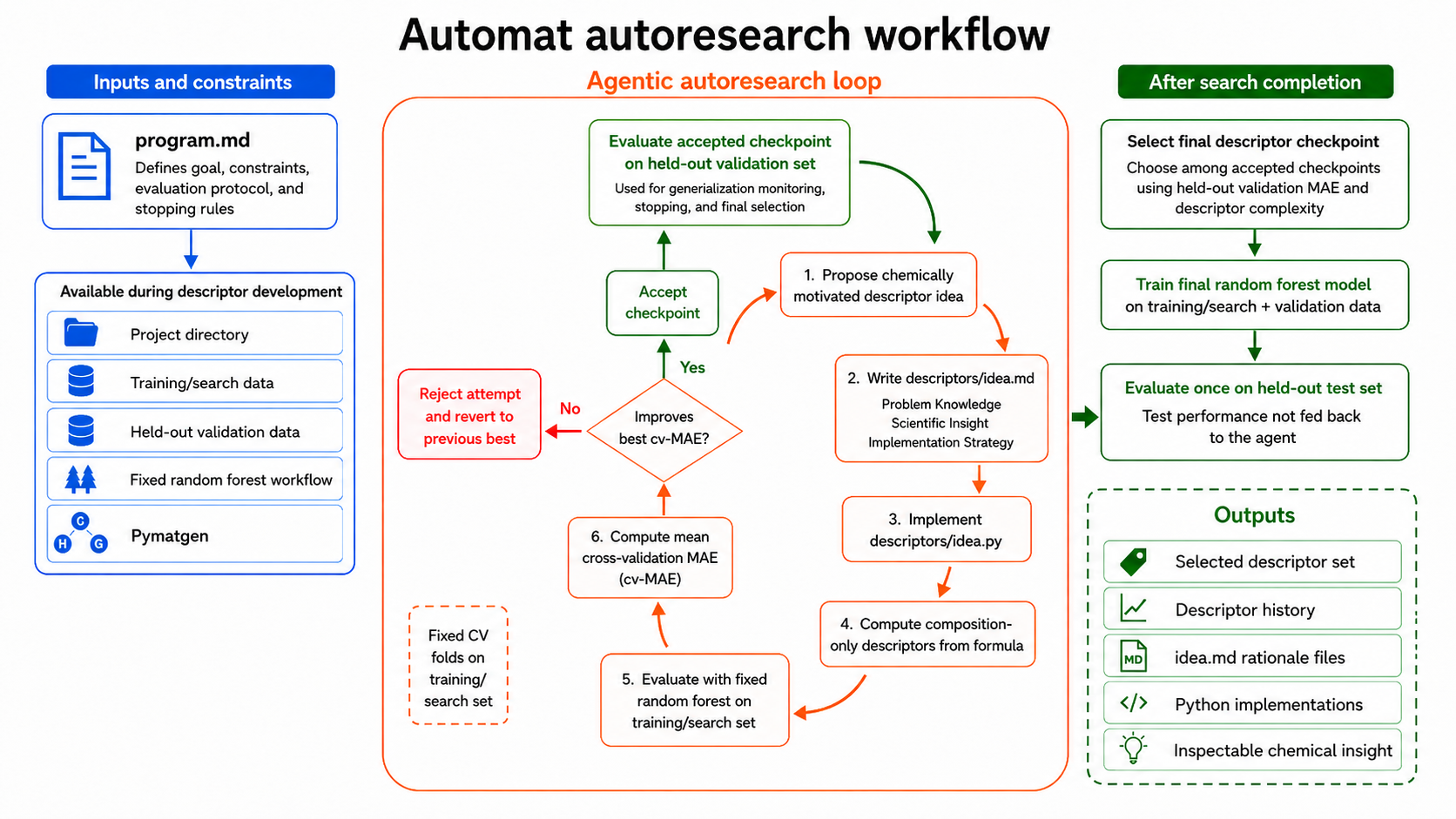}
    \caption{
    Schematic representation of the {\sc Automat} autoresearch workflow defined in \texttt{program.md}. The agent is given 
    access to the project directory, the training/search data, and the held-out validation data. However, the held-out test 
    set is not used during the descriptor development. At each iteration, the agent proposes a chemically motivated 
    composition-only descriptor strategy, records its rationale and implementation plan in \texttt{descriptors/idea.md}, 
    implements the corresponding Python code in \texttt{descriptors/idea.py}, and evaluates the resulting descriptors 
    using a random forest model. Local accept/reject decisions are based on the mean cross-validation MAE computed 
    on fixed folds of the training/search set: descriptor modifications that improve this inner cross-validation metric are 
    retained, whereas unsuccessful attempts are discarded and the workflow reverts to the previous best checkpoint. 
    Whenever a descriptor update is accepted, the updated representation is also evaluated on the held-out validation set. 
    This held-out validation performance is not used for local accept/reject decisions, but only to monitor generalization, 
    to apply stopping criteria, and to select the final descriptor checkpoint, while also considering descriptor complexity. 
    After the autoresearch run is complete, a final random forest model is trained using the selected descriptor set on 
    the combined training/search and validation data and evaluated once on the held-out test set. The accumulated descriptor 
    history, rationale files, and implementations provide an auditable record of the descriptor-design trajectory and can be 
    inspected to identify chemically meaningful feature families.
    }
    \label{workflow}
\end{figure*}

\section{Introduction}

The discovery of materials with technologically relevant properties remains a central challenge in materials 
science~\cite{curtarolo2013_highthroughput, butler2018_ml, schmidt2019_solidstate}. As established technologies 
approach their performance limits, there is an increasing demand for novel compounds capable of enabling new 
or improved functionality. Machine learning provides a powerful route to accelerate this process by learning from 
existing experimental data and by using this knowledge to prioritize candidate materials before resources are 
committed to synthesis and characterization.

Composition-based models are particularly attractive in this context, since they require only chemical formulas as input. By training on experimental data, such models can predict the properties of previously unexplored compounds without requiring crystallographic or structural information, which is typically unavailable or difficult to extract, especially when the measured property and the crystal structure must correspond to the same sample. Composition-based approaches have 
achieved strong performance across a range of materials-property prediction tasks~\cite{magpie, James_tc, roost, curtarolo_tc}. 
However, their success depends critically on how chemical formulas are represented as numerical inputs. Although 
this problem has been extensively studied~\cite{magpie, James_tc, roost, crabnet, matbench, oganov_tc}, 
selecting an effective representation remains nontrivial, task-dependent, and often reliant on substantial domain 
expertise.

The dependence on representation is particularly sensitive in low-data regimes~\cite{zhang2018_small,xu2023_small}. 
Unfortunately, many experimental materials datasets are small compared to those typically used in large-scale machine 
learning, and much of the relevant knowledge in materials science remains embedded in an unstructured form
in the scientific literature, including text, tables, and figures. This information is difficult to extract and, despite many attempts~\cite{olivetti2020_nlp, ChemDataExtractor, bert_psie, chatextract}, comprehensive databases of 
experimental data remain very limited. Consequently, supervised models trained on curated experimental datasets 
often operate with limited data. In such settings, it is difficult to rely solely on large models to learn rich representations 
directly from the training set. Instead, descriptors must expose chemically and physically relevant information in a form 
that can be exploited by the learning algorithm.

Early composition-based models often combined relatively simple learning algorithms, such as random 
forests~\cite{random_forests}, with carefully engineered descriptors derived from chemical 
composition~\cite{magpie, James_tc, modnet1, curtarolo_tc}. These descriptors transform a chemical formula 
into a numerical vector, commonly by computing statistics or composition-weighted averages of elemental properties 
over the constituent atoms~\cite{magpie, matminer}. The choice of elemental properties, aggregation rules, and 
mathematical transformations is typically guided by chemical intuition and refined through empirical validation. 
The success of these approaches demonstrates that, in small experimental datasets, predictive performance can 
depend as strongly on descriptor quality as on the learning algorithm itself.

More recent approaches based on attention mechanisms, graph representations, and large language models have 
expanded the range of architectures available for materials-property prediction~\cite{crabnet, gptchem, darwin}. 
These models can benefit from pre-training or broader exposure to scientific text and have produced important 
advances. Nevertheless, general-purpose models may exhibit inconsistent performance across 
tasks~\cite{gptchem, darwin, matbench}, often require fine-tuning, and can still be outperformed by smaller, task-specific 
models equipped with carefully designed input features. Manual descriptor design therefore remains valuable, but it is also 
limiting: it requires domain expertise, extensive experimentation, and significant researcher time. Furthermore, 
descriptors optimized for one property or dataset may not transfer effectively to another.

This bottleneck is becoming increasingly important as large language model workflows make it more feasible to 
extract structured materials data from the scientific literature~\cite{ChemDataExtractor, bert_psie, nemad}. With 
new experimental datasets becoming available, new methods are needed that can rapidly develop high-performing, 
task-specific models without requiring extensive manual feature engineering for each new problem.
Existing automated workflows in materials informatics have primarily focused on constructing predictive pipelines, 
including featurization, feature reduction, model selection, and hyperparameter optimization~\cite{matbench, modnet1, modnet2}. 
Automated descriptor discovery has also been explored, most notably through symbolic regression and compressed-sensing 
approaches such as SISSO, which searches for compact and interpretable relationships among predefined physical or 
chemical quantities~\cite{sisso,symbolic_regression}. However, these methods generally operate within human-specified 
feature spaces, operator sets, and transformation rules.

Recent progress in large language models has enabled agentic systems that combine language-based reasoning with the
use of tools, memory, and interaction with external computational 
environments~\cite{bran2024_chemcrow,boiko2023_coscientist,yang2024_sweagent, el_agente}. This development 
has been especially prominent in code-oriented applications, where autonomous or semi-autonomous agents can generate 
programs, modify existing codebases, run tests, diagnose failures, and improve solutions through iterative feedback. These 
capabilities suggest a broader role for code agents in scientific research. Such agents can participate directly in executable 
research workflows by proposing computational strategies, implementing them, evaluating their outcomes, and refining 
subsequent attempts. This makes agent-based approaches particularly attractive for descriptor design, where scientific reasoning 
must be coupled to reproducible numerical evaluation.

In this work, we investigate whether autonomous research agents can be used to design compositional descriptors for 
materials-property prediction. We build on the autoresearch paradigm introduced by Andrej Karpathy~\cite{autoresearch}, 
in which an autonomous agent iteratively proposes, implements, evaluates, and refines research ideas to improve the training 
of a large language model. Rather than using autonomous research primarily to optimize model architectures or hyperparameters, 
we focus on the design of model input features. This task is deliberately constrained but scientifically meaningful: effective 
descriptor design requires identifying chemical and physical quantities relevant to a target property and expressing them in a 
form suitable for machine learning.

We introduce {\sc Automat}, an autoresearch framework for designing compositional descriptors for chemical compounds. {\sc Automat} 
evaluates generated descriptors within a reproducible random forest modeling protocol. At each iteration, the agent proposes 
descriptor-generation strategies, implements them, evaluates their predictive performance, and uses the results to guide subsequent 
proposals. The framework is configured to encourage scientifically motivated descriptors rather than arbitrary numerical transformations 
of the input formula. The goal of this study is not necessarily to outperform all state-of-the-art materials-property predictors. Instead, 
we use a deliberately fixed modeling workflow to isolate a specific question: whether autonomous descriptor design can improve 
predictive performance when the learning algorithm and evaluation protocol are held constant. This setup provides both a practical 
platform for evaluating autonomous agentic descriptor design in materials modeling and an extensible baseline for researchers 
developing related autoresearch workflows.

We show that {\sc Automat} can autonomously generate descriptors for random forest models that improve upon fixed-workflow 
random forest baselines. Beyond predictive accuracy, we demonstrate that the outputs of an autonomous descriptor-design run 
can provide insight into the underlying modeling problem, including relevant elemental properties, chemical trends, and task-specific 
descriptor families. These results suggest that autonomous descriptor design can support materials researchers and R\&D 
practitioners in understanding datasets, identifying informative chemical features, and accelerating the development of task-specific 
materials models.

\section{Methods}

In this work, we adapt the autoresearch workflow introduced by Karpathy~\cite{autoresearch} to the problem of 
designing compositional descriptors for materials-property prediction. Rather than using the agentic loop to optimize 
model architectures or hyperparameters, {\sc Automat} limits the search to the construction of numerical descriptors 
from chemical formulas. All predictive models use a random forest architecture, and each autoresearch iteration proposes, 
implements, evaluates, and either accepts or rejects a candidate descriptor set. Random forests are well suited to tabular 
compositional descriptors, are robust in low-data regimes, and have been widely used in materials-informatics 
baselines~\cite{random_forests, magpie, James_tc, curtarolo_tc}.

The central premise of the autoresearch paradigm is that modern LLM-controlled coding agents can interact with a 
computational environment in a sufficiently reliable way to support iterative scientific workflows~\cite{react, yang2024_sweagent}. 
In this setting, the agent is provided with written instructions, access to the project directory, and the ability to write, execute, 
debug, and modify code. The agent receives numerical feedback from an evaluation environment and uses this feedback 
to guide subsequent iterations.

{\sc Automat} applies this paradigm specifically to descriptor design. At each iteration, the agent proposes a chemically 
motivated descriptor-generation strategy, implements it as executable Python code, evaluates the resulting descriptors 
within a fixed random forest workflow, and determines whether the new descriptor set improves performance relative to 
the current best checkpoint. The objective is not to search over arbitrary numerical transformations, but to construct descriptors 
that encode physically and chemically meaningful information from the input composition.

To keep the descriptor-design task well defined, the agent is restricted to information available from the chemical formula. 
Pymatgen~\cite{pymatgen} is used for chemical-composition parsing and elemental-property information. No structural information, 
external materials databases, or test-set labels are available to the agent during descriptor development.

\subsection{The autoresearch loop: \texttt{program.md}}

The autoresearch loop is specified in a \texttt{program.md} file. This is the first file that the agent is instructed to read at the 
beginning of each run. The file defines the goal of the experiment, the constraints on the descriptor generation, the available 
computational resources, the evaluation protocol, and the rules governing acceptance or rejection of candidate descriptor sets. 
A schematic representation of the workflow is shown in Fig.~\ref{workflow}.  For each target task, {\sc Automat} requires three 
non-overlapping datasets: a training/search set, a held-out validation set, and a held-out test set. The test labels are never made 
available to the agent during the descriptor development. The random forest hyperparameters are specified at the beginning of 
the run and kept fixed throughout the descriptor-search procedure. This ensures that changes in performance can be attributed 
primarily to descriptor design rather than to model tuning.

At the beginning of each iteration, the agent proposes a new descriptor strategy and implements it in Python. The descriptors 
are computed from the chemical formulas and used as input features for a random forest model implemented with 
scikit-learn~\cite{scikit-learn}. Candidate descriptor sets are evaluated using the validation protocol described below. If a candidate 
improves the current optimization metric, it is accepted and becomes the new reference checkpoint for subsequent iterations. 
Otherwise, it is discarded and the search resumes from the previous best checkpoint. This accept/reject mechanism provides a 
simple hill-climbing procedure over descriptor space.

\subsection{Two-level validation protocol}

The descriptor optimization is performed using a nested validation strategy designed to reduce overfitting during the search 
while preserving an unbiased final test set. The training/search set is used for descriptor generation and cross-validation-based 
update decisions. The held-out validation set is used as an outer model-selection and stopping criterion during descriptor search. 
The held-out test set is reserved exclusively for final evaluation.

At the start of each run, the training/search set is partitioned into a fixed stratified $n$-fold cross-validation split. For regression 
tasks, stratification is performed by binning the target variable before fold assignment. This split is generated once and kept 
unchanged throughout all descriptor-search iterations. Keeping the folds fixed avoids stochastic fluctuations in the cross-validation 
score that could otherwise lead to spurious descriptor acceptance or rejection.
At each iteration, a candidate descriptor set is evaluated using random forest regression. For each fold, a random forest regressor is trained on the corresponding cross-validation training partition and evaluated on the associated validation fold. The mean absolute error (MAE), averaged across all validation folds, is used as the optimization signal, denoted as cv-MAE. A candidate descriptor set is accepted only if it improves the cv-MAE relative to the current best descriptor set.

Whenever a descriptor update is accepted according to the cv-MAE criterion, an additional random forest model is trained on the 
full training/search set using the accepted descriptor set and evaluated on the held-out validation set. The resulting validation MAE 
is used to monitor whether improvements in cv-MAE translate to improved generalization beyond the cross-validation folds. This 
metric also serves as a stopping criterion: if the held-out validation MAE fails to improve over a predefined number of accepted 
descriptor updates, subsequent reductions in cv-MAE are treated as likely evidence of overfitting to the training/search set.

At the end of the autoresearch run, the final descriptor set is therefore not selected solely on the basis of the lowest cv-MAE. 
Instead, the selected descriptor set is chosen from among the descriptor sets accepted during the search by considering both 
held-out validation performance and descriptor complexity. Priority is given to descriptor sets that achieve the lowest held-out 
validation MAE, with lower-complexity representations preferred when validation performance is comparable. In this way, the 
validation set acts as an outer model-selection criterion while remaining separate from the cross-validation folds used for iterative 
descriptor acceptance, and the final selection favours descriptors that generalize well without unnecessary feature complexity.

After the descriptor search is complete, a final random forest model is trained using the selected descriptor set on the combined 
training/search and validation sets. This model is evaluated once on the held-out test set, which was inaccessible to the agent 
throughout the autoresearch run. The test-set performance is not fed back to the agent and is not used to guide subsequent 
descriptor design. This procedure prevents leakage from the test set into the descriptor-design process and preserves the test 
set as an unbiased estimate of final model performance.

The optimized descriptors are compared against baseline descriptor representations using the same final evaluation protocol. 
Since the model architecture and final training procedure are fixed across all descriptor sets, the comparison isolates the contribution 
of the automated descriptor design.

\subsection{Planning strategy: \texttt{idea.md}}

Since the descriptor design requires scientific judgment as well as numerical evaluation, {\sc Automat} requires the agent to 
document each proposed descriptor strategy before implementation. At every iteration, the agent first writes a natural-language 
plan in \texttt{descriptors/idea.md}. Only after this file has been updated can the agent implement the corresponding descriptor 
code in \texttt{descriptors/idea.py}.

Each \texttt{idea.md} file contains three required sections:

\begin{itemize}
    \item \textbf{Problem Knowledge}: a concise summary of the target task and relevant observations 
    accumulated from previous iterations.
    \item \textbf{Scientific Insight}: the chemical or physical reasoning motivating the proposed descriptor 
    strategy.
    \item \textbf{Implementation Strategy}: a description of the features to be implemented and a rationale for why they are expected to improve the model performance.
\end{itemize}

This documentation step encourages the agent to formulate descriptor proposals as interpretable chemical 
or physical hypotheses rather than arbitrary feature transformations, while also enabling interrupted runs to 
be restarted by a user or another agent without requiring access to the full prior conversation history.

The \texttt{idea.md} file therefore provides an inspectable record of the reasoning behind each descriptor 
proposal. Since an agent's internal reasoning trace is often not exposed by the LLM provider and cannot 
serve as part of the scientific record, {\sc Automat} requires the relevant scientific rationale to be externalized 
in a persistent file that can be audited after the run.

\subsection{Descriptor implementation}

The executable descriptor logic is stored in \texttt{descriptors/idea.py}. This file defines the transformation 
from a chemical formula to a fixed-length numerical feature vector suitable for use in a scikit-learn random 
forest model. Each descriptor function receives a chemical composition as input, parses it using Pymatgen, 
and returns numerical descriptors derived only from the composition.

The agent is allowed to construct descriptors from quantities available through Pymatgen, including elemental 
properties, stoichiometric information, oxidation-state information, and composition-derived statistics. Typical 
operations include composition-weighted averages, extrema, ranges, variances, element fractions, and 
chemically motivated combinations of elemental quantities.

Care is taken to ensure that the {\sc Automat} repository does not contain information that could bias the generated 
descriptors toward a particular design. During each run, the agent therefore relies only on its pretrained knowledge, 
the information contained in the local project files, and the numerical feedback obtained from previous descriptor 
evaluations.

The descriptor implementations must be deterministic and robust to invalid or unusual formulas. If a descriptor 
calculation fails, returns non-finite values, or exceeds the allowed runtime, the iteration will be treated as 
unsuccessful. The agent must then diagnose the failure, revert to the previous best descriptor checkpoint, and 
continue the autoresearch loop from that state.

\subsection{Running an {\sc Automat} experiment}

An {\sc Automat} run can be launched from an agentic coding interface such as OpenAI Codex or 
Claude Code~\cite{openai-codex, claude-code}. At the beginning of a new experiment, the user instructs the agent 
to initialize the run from the instructions in \texttt{program.md}. For example, a new run can be started with the following 
prompt:

\begin{refbox}
Set up a new experiment run. Follow strictly the directives in program.md.
\end{refbox}

The agent then creates a new git branch, initializes the experiment directory, executes a first baseline evaluation, and 
stops for user input. The autoresearch loop can then be initiated with the following prompt:

\begin{refbox}
Continue performing new iterations, strictly following the instructions in program.md. Continue until run\_status.py says STOP.
\end{refbox}

The stopping logic is implemented in a \texttt{run\_status.py} script so that the agent does not need to track complex halting 
conditions internally. The user may also specify alternative stopping criteria or provide high-level feedback to guide the search. 
In the default implementation, the run stops either when the maximum number of iterations, $N_{\mathrm{max}}$, is reached or 
when the held-out validation MAE has not improved for a predefined number of accepted descriptor updates, 
denoted as \texttt{n\_validation\_patience}. In all cases, the agent is instructed to follow the constraints defined in 
\texttt{program.md}, including the use of composition-only descriptors, Pymatgen-based descriptor construction and a fixed 
random forest evaluation workflow.

This setup allows {\sc Automat} to be used either as a fully autonomous descriptor-design system or as a collaborative 
research assistant. In the latter mode, a human researcher can periodically review the generated ideas, inspect descriptor 
implementations, and provide high-level guidance for subsequent iterations.


\begin{figure*}[]
    \centering 
    \includegraphics[width=1\linewidth]{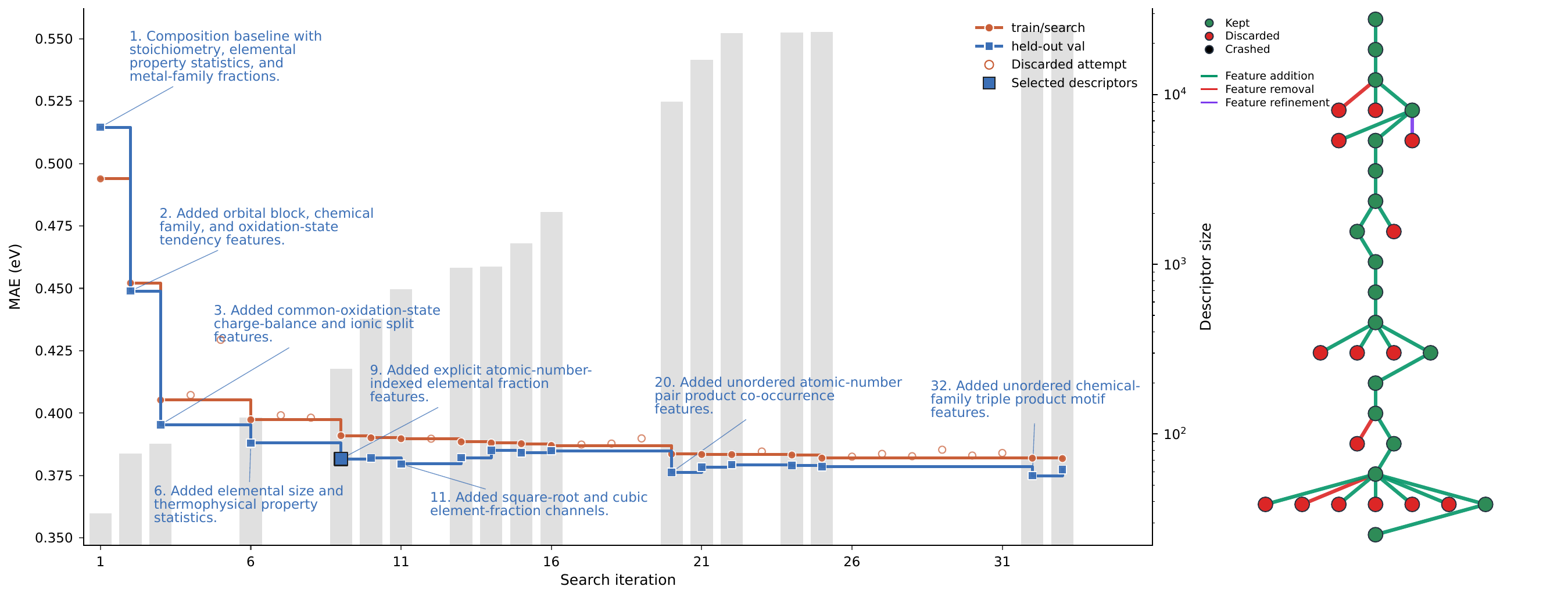}
    \caption{
    {\sc Automat} descriptor-design trajectory for composition-only prediction of experimental band gaps. 
    Left panel: model performance and descriptor dimensionality as functions of the autoresearch iteration. The orange curve 
    shows the running-best MAE on the training/search cross-validation folds, which is used by the agent for local 
    accept/reject decisions. The blue curve shows the MAE on the held-out validation set for accepted descriptor checkpoints, 
    which is used for outer model selection. Filled markers indicate accepted descriptor modifications, open markers indicate 
    discarded attempts, and the larger square marks the descriptor set selected for final test evaluation. Gray bars show the 
    dimensionality of the descriptor set at accepted checkpoints (right-hand side log scale). Text annotations identify the main 
    descriptor families introduced during the run. 
    Right panel: graphical representation of the same descriptor-search trajectory. Green nodes denote accepted descriptor sets, 
    while red nodes denote discarded attempts. Edge colors indicate whether the corresponding agent action added descriptors, 
    removed descriptors, or refined an existing descriptor family. The largest validation improvement occurs early in the run, after 
    the introduction of oxidation-state, charge-balance, and ionic-splitting descriptors. Later feature expansions reduce the 
    cross-validation MAE but substantially increase descriptor dimensionality and do not improve the selected held-out-validation 
    trade-off. Descriptor families in the selected representation are summarized in Table~\ref{tab:gap}. For visualization purposes, 
    rejected descriptor updates occurring after the final accepted iteration are omitted.
    }
    \label{bandgaprun}
\end{figure*}


\begin{table*}[t]
\centering
\caption{Descriptor families generated by {\sc Automat} for composition-only prediction of experimental band gaps. The selected 
243-dimensional representation is derived exclusively from the reduced chemical formula and combines stoichiometric statistics, 
composition-weighted elemental-property summaries, element-family fractions, oxidation-state and ionic-balance terms, size and 
thermodynamic descriptors, radius-contrast terms, and an element-wise fractional composition array. Here, $i$ indexes elements 
in the composition, $x_i$ is the atomic fraction of element $i$, $n_i$ is its reduced-formula stoichiometric coefficient, $p_i$ is a 
generic elemental property, $q_i$ is a candidate oxidation state, $r_i$ is an atomic radius, and $T_i$ is a temperature-related 
elemental property such as the melting or boiling point.}
\label{tab:gap}
\small
\setlength{\tabcolsep}{0pt}
\renewcommand{\arraystretch}{1.15}

\begin{tabular*}{\textwidth}{@{\extracolsep{\fill}}lrll@{}}
\textbf{Type} & \textbf{Size} & \textbf{Description} & \textbf{Example} \\
\midrule
\midrule

\parbox[t]{0.20\textwidth}{\raggedright Stoichiometric descriptors} &
6 &
\parbox[t]{0.42\textwidth}{\raggedright Counts, fraction extrema, and composition diversity from the reduced formula} &
\parbox[t]{0.24\textwidth}{\raggedright
$N, x_{\max}, x_{\min}, -\sum_i x_i \log x_i$} \\

\midrule

\parbox[t]{0.20\textwidth}{\raggedright Weighted elemental properties} &
25 &
\parbox[t]{0.42\textwidth}{\raggedright Composition-weighted means, extrema, ranges, and spreads of basic elemental attributes} &
\parbox[t]{0.24\textwidth}{\raggedright
$\bar{p}, \max_i p_i, \min_i p_i, \sigma(p)$} \\

\midrule

\parbox[t]{0.20\textwidth}{\raggedright Element family fractions} &
16 &
\parbox[t]{0.42\textwidth}{\raggedright Atomic fractions grouped by metallic character, orbital block, and periodic-table family} &
\parbox[t]{0.24\textwidth}{\raggedright
$\sum_{i \in d} x_i, \sum_{i \in \mathrm{halogen}} x_i$} \\

\midrule

\parbox[t]{0.20\textwidth}{\raggedright Oxidation-state descriptors} &
30 &
\parbox[t]{0.42\textwidth}{\raggedright Candidate oxidation-state ranges, counts, polarity flags, and weighted spreads} &
\parbox[t]{0.24\textwidth}{\raggedright
$\min_i q_i, \max_i q_i, \sigma(q)$} \\

\midrule

\parbox[t]{0.20\textwidth}{\raggedright Ionic-balance descriptors} &
11 &
\parbox[t]{0.42\textwidth}{\raggedright Best-charge-assignment residuals, neutral-solution prevalence, ionic strength, and cation-anion partitioning} &
\parbox[t]{0.24\textwidth}{\raggedright
$\lvert \sum_i n_i q_i \rvert / \sum_i n_i, \sum_i x_i \lvert q_i \rvert$} \\

\midrule

\parbox[t]{0.20\textwidth}{\raggedright Size and thermodynamic properties} &
35 &
\parbox[t]{0.42\textwidth}{\raggedright Weighted statistics of radius, volume, density, melting point, and boiling point attributes} &
\parbox[t]{0.24\textwidth}{\raggedright
$\sum_i x_i r_i, \max_i T_i - \min_i T_i$} \\

\midrule

\parbox[t]{0.20\textwidth}{\raggedright Radius contrast descriptors} &
2 &
\parbox[t]{0.42\textwidth}{\raggedright Derived radius scale and mismatch summaries from constituent atomic radii} &
\parbox[t]{0.24\textwidth}{\raggedright
$\exp(\sum_i x_i \log r_i), r_{\max}/r_{\min}$} \\

\midrule

\parbox[t]{0.20\textwidth}{\raggedright Fractional composition array} &
118 &
\parbox[t]{0.42\textwidth}{\raggedright Element-wise composition entries indexed by atomic number from hydrogen through oganesson} &
\parbox[t]{0.24\textwidth}{\raggedright
$x_{\mathrm{H}}, x_{\mathrm{O}}, x_{\mathrm{Fe}}$} \\

\midrule
\midrule

\parbox[t]{0.20\textwidth}{\raggedright Total} &
243 &
\parbox[t]{0.42\textwidth}{\raggedright} &
\parbox[t]{0.24\textwidth}{} \\

\end{tabular*}
\end{table*}


\begin{figure*}[]
    \centering 
    \includegraphics[width=1\linewidth]{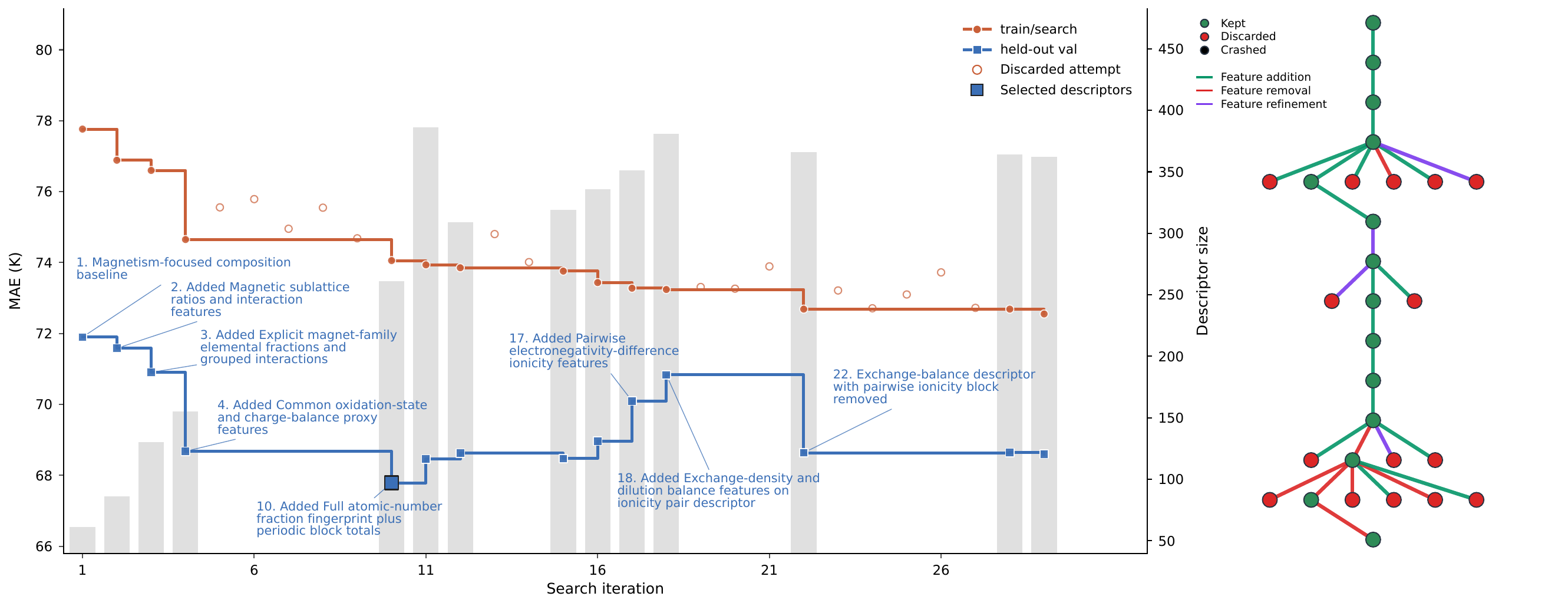}
    \caption{
    {\sc Automat} descriptor-design trajectory for composition-only prediction of the experimental Curie temperature, $T_\mathrm{C}$, of 
    permanent ferromagnets. 
    Left panel: model performance and descriptor dimensionality as functions of the autoresearch iteration. The orange curve shows the running-best 
    MAE on the training/search cross-validation folds, which is used by the agent for local accept/reject decisions. The blue curve shows the 
    MAE on the held-out validation set for accepted descriptor checkpoints, which is used for outer model selection. Filled markers indicate 
    accepted descriptor modifications, open markers indicate discarded attempts, and the larger square marks the descriptor set selected 
    for final test evaluation. Gray bars show the dimensionality of the descriptor set at accepted checkpoints  (right-hand side scale). 
    Text annotations identify the main chemically motivated descriptor families introduced during the run. 
    Right panel: graphical representation of the same descriptor-search trajectory. Green nodes denote accepted descriptor sets, and 
    red nodes denote discarded attempts. Edge colors indicate whether the corresponding agent action added descriptors, removed 
    descriptors, or refined an existing descriptor family. The selected checkpoint occurs at iteration 10, where the held-out validation 
    MAE reaches its minimum before later accepted updates reduce the cross-validation MAE without improving held-out validation 
    performance. Descriptor families in the selected representation are summarized in Table~\ref{tab:tc}. For visualization purposes, 
    rejected descriptor updates occurring after the final accepted iteration are omitted.
    }
    \label{tcrun}
\end{figure*}

\begin{table*}[t]
\centering
\caption{Descriptor families generated by {\sc Automat} for composition-only prediction of experimental Curie temperatures 
of permanent ferromagnets. The selected 261-dimensional representation is derived exclusively from the reduced chemical 
formula and combines stoichiometric statistics, targeted magnetic-chemistry descriptors, elemental-property summaries, 
magnetic-sublattice and family-interaction terms, valence-balance descriptors, periodic-block terms, and an atomic-number-indexed 
fractional composition array. Here, $i$ indexes elements in the composition, $x_i$ is the atomic fraction of element $i$, $n$ is 
the number of distinct elements, $p_i$ is a generic elemental property, and $q_i^{+}$ and $q_i^{-}$ are representative positive 
and negative oxidation-state priors. Grouped quantities such as $x_{\mathrm{TM}}$, $x_{\mathrm{RE}}$, $x_{\mathrm{Act}}$, 
$x_{\mathrm{FeCo}}$, $x_d$, and $x_f$ denote summed atomic fractions over transition-metal, rare-earth, actinide, Fe/Co, 
$d$-block, and $f$-block elements, respectively.}
\label{tab:tc}
\small
\setlength{\tabcolsep}{0pt}
\renewcommand{\arraystretch}{1.15}

\begin{tabular*}{\textwidth}{@{\extracolsep{\fill}}lrll@{}}
\textbf{Type} & \textbf{Size} & \textbf{Description} & \textbf{Example} \\
\midrule
\midrule

\parbox[t]{0.20\textwidth}{\raggedright Stoichiometric descriptors} &
5 &
\parbox[t]{0.42\textwidth}{\raggedright Quantities derived from composition size, diversity, and concentration} &
\parbox[t]{0.24\textwidth}{\raggedright
$n, -\sum_i x_i \log x_i, \max_i x_i$} \\

\midrule

\parbox[t]{0.20\textwidth}{\raggedright Targeted chemistry descriptors} &
21 &
\parbox[t]{0.42\textwidth}{\raggedright Fractions, ratios, and products for magnetic, metallic, heavy-element, and anion families} &
\parbox[t]{0.24\textwidth}{\raggedright
$x_{\mathrm{Fe}}, \sum_{i \in \mathrm{3d}} x_i, x_{\mathrm{TM}} \cdot x_{\mathrm{O}}$} \\

\midrule

\parbox[t]{0.20\textwidth}{\raggedright Weighted elemental properties} &
7 &
\parbox[t]{0.42\textwidth}{\raggedright Composition-weighted summaries of constituent attributes} &
\parbox[t]{0.24\textwidth}{\raggedright
$\sum_i x_i p_i$} \\

\midrule

\parbox[t]{0.20\textwidth}{\raggedright Property extrema} &
14 &
\parbox[t]{0.42\textwidth}{\raggedright Minimum and maximum elemental attributes across constituents} &
\parbox[t]{0.24\textwidth}{\raggedright
$\min_i p_i, \max_i p_i$} \\

\midrule

\parbox[t]{0.20\textwidth}{\raggedright Property spread descriptors} &
14 &
\parbox[t]{0.42\textwidth}{\raggedright Variability and range of elemental attributes under composition weights} &
\parbox[t]{0.24\textwidth}{\raggedright
$\sigma(p), \max_i p_i - \min_i p_i$} \\

\midrule

\parbox[t]{0.20\textwidth}{\raggedright Magnetic sublattice descriptors} &
25 &
\parbox[t]{0.42\textwidth}{\raggedright Fractions, ratios, contrasts, and products for magnetic 3d, rare-earth, actinide, and anion chemistry} &
\parbox[t]{0.24\textwidth}{\raggedright
$x_{\mathrm{FeCo}} / x_{\mathrm{TM}}, x_{\mathrm{FeCo}} \cdot x_{\mathrm{REAct}}$} \\

\midrule

\parbox[t]{0.20\textwidth}{\raggedright Family identity fractions} &
35 &
\parbox[t]{0.42\textwidth}{\raggedright Element-wise and grouped fractions for selected rare-earth, actinide, pnictogen, interstitial, and main-group families} &
\parbox[t]{0.24\textwidth}{\raggedright
$x_{\mathrm{Nd}}, \sum_{i \in \mathrm{RE}} x_i, \sum_{i \in \mathrm{Act}} x_i$} \\

\midrule

\parbox[t]{0.20\textwidth}{\raggedright Family interaction terms} &
9 &
\parbox[t]{0.42\textwidth}{\raggedright Products coupling selected family fractions with magnetic or transition-metal fractions} &
\parbox[t]{0.24\textwidth}{\raggedright
$x_{\mathrm{NdSmY}} \cdot x_{\mathrm{FeCo}}$} \\

\midrule

\parbox[t]{0.20\textwidth}{\raggedright Valence prior summaries} &
12 &
\parbox[t]{0.42\textwidth}{\raggedright Weighted means and spreads of common oxidation-state priors} &
\parbox[t]{0.24\textwidth}{\raggedright
$\sum_i x_i q_i^{+}, \sigma(q^{-})$} \\

\midrule

\parbox[t]{0.20\textwidth}{\raggedright Valence balance terms} &
13 &
\parbox[t]{0.42\textwidth}{\raggedright Oxidation-capacity balances, ratios, spans, and chemistry-coupled capacities} &
\parbox[t]{0.24\textwidth}{\raggedright
$\sum_i x_i q_i^{+} - \sum_i x_i q_i^{-}$} \\

\midrule

\parbox[t]{0.20\textwidth}{\raggedright Fractional composition array} &
96 &
\parbox[t]{0.42\textwidth}{\raggedright Fixed atomic-number-indexed fractions from hydrogen through curium} &
\parbox[t]{0.24\textwidth}{\raggedright
$x_Z$} \\

\midrule

\parbox[t]{0.20\textwidth}{\raggedright Periodic block terms} &
10 &
\parbox[t]{0.42\textwidth}{\raggedright Broad periodic-block fractions plus block, anion, and magnetic interaction ratios} &
\parbox[t]{0.24\textwidth}{\raggedright
$x_d \cdot x_f, x_{\mathrm{FeCo}} / x_d$} \\

\midrule
\midrule

\parbox[t]{0.20\textwidth}{\raggedright Total} &
261 &
\parbox[t]{0.42\textwidth}{\raggedright} &
\parbox[t]{0.24\textwidth}{} \\

\end{tabular*}
\end{table*}

\begin{figure*}[]
    \centering 
    \includegraphics[width=1\linewidth]{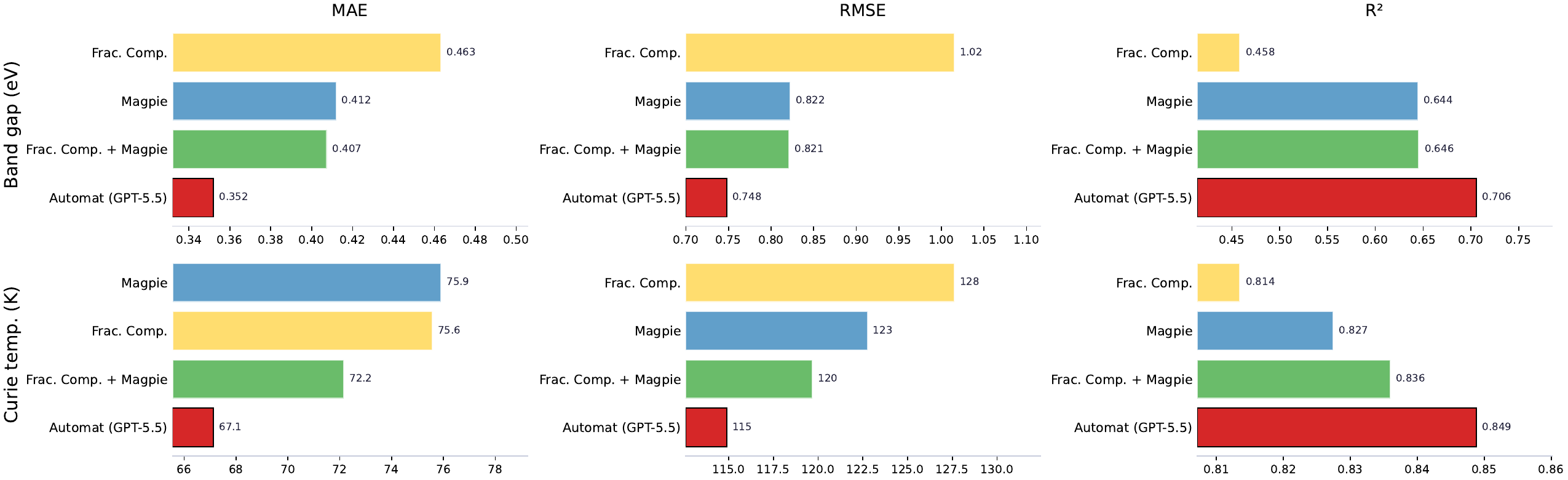}    
    \caption{
    Final held-out test performance of random forest models using different composition-only descriptor representations. Results 
    are shown for experimental band-gap prediction in eV and Curie-temperature prediction in K. Columns report the MAE, root 
    mean squared error (RMSE), and coefficient of determination ($R^2$). Lower values are better for MAE and RMSE, whereas 
    higher values are better for $R^2$, which is upper-bounded to 1. Baselines include a fractional composition array indexed by atomic 
    number (118 components), Magpie descriptors (132 components), and the concatenation of fractional composition and Magpie descriptors (250 components). {\sc Automat} denotes 
    the descriptor set generated by the autoresearch workflow using GPT-5.5 in the OpenAI Codex harness. All models use the 
    same random forest architecture and the same final training/validation/test protocol, so differences in performance reflect the 
    effect of the descriptor representation. {\sc Automat} improves over all baseline descriptor sets for both target properties.
    }
    \label{benchmark}
\end{figure*}

\section{Results}

We evaluate {\sc Automat} on two composition-only regression tasks: the prediction of experimental band gaps (\texttt{expt\_gap}) 
and that of the experimental Curie temperatures of permanent ferromagnets (\texttt{expt\_Tc}). In both tasks, the only model 
input is the chemical formula. Atomic structures are not used, allowing us to isolate compositional descriptor design from the 
separate problem of representing atomic coordinates for machine learning~\cite{behlerparrinello, bispectrum, ace, MTP}.

For each task, a budget of 50 autoresearch iterations is allocated to {\sc Automat}. At each iteration, the agent proposes 
a chemically motivated modification to the descriptor set, implements the corresponding Python code, evaluates the resulting 
representation using the random forest workflow described in the Methods section, and retains the modification only if it 
improves the three-fold cv-MAE on the training/search set. Whenever a descriptor update is accepted, the updated representation 
is also evaluated on a held-out validation set. This held-out validation performance is never used for the local accept/reject decision, 
but only to monitor generalization during the run and to select the final descriptor set.

Because autoresearch is still an emerging paradigm, there is not yet a standard format for reporting agent-generated scientific 
workflows. We therefore report both the final predictive performance and the descriptor-design trajectories. In particular, we 
analyze the descriptor families introduced by {\sc Automat}, the chemically meaningful changes associated with performance 
improvements, and the limitations observed during the runs.

All experiments were performed using the {\sc Automat} autoresearch workflow with GPT-5.5 as the coding agent in the OpenAI Codex harness at medium reasoning effort~\cite{openai, openai-codex}. To avoid task-specific prompt engineering, 
the initial task description supplied to the agent was deliberately minimal. For the two tasks considered here, the problem 
statements were limited to one-line descriptions: ``Predict the experimental band gap of inorganic materials from chemical formula only'' and 
``Predict the experimental Curie temperature of ferromagnets from chemical formula only.''

We compare the descriptors generated by {\sc Automat} against random forest models trained with three baseline descriptor sets:

\begin{itemize}
    \item \textbf{Fractional composition array:} each composition is represented by a fixed-length vector indexed by atomic 
    number. The value at each element index is the atomic fraction of that element in the composition, with absent elements 
    set to zero. This descriptor contains no elemental-property information and therefore requires the model to infer chemical 
    similarity entirely from the training data.

    \item \textbf{Magpie:} elemental-property descriptors as implemented in {\sc matminer}~\cite{magpie, matminer}. For 
    each composition, tabulated elemental properties, including atomic number, atomic weight, Mendeleev number, melting 
    temperature, covalent radius, electronegativity, and valence-electron counts, are aggregated over the constituent elements 
    using summary statistics such as minimum, maximum, range, mean, mean absolute deviation, and mode.

    \item \textbf{Fractional composition array + Magpie:} the concatenation of the fractional composition array and the Magpie 
    descriptor vector.
\end{itemize}

We compare {\sc Automat} with the baseline descriptor sets using the same random forest model and evaluation protocol. 
This comparison is designed to test whether the agent-generated descriptors improve the representation of chemical composition, 
rather than whether the overall workflow establishes a new state-of-the-art predictor. The final held-out test performance is summarized 
in Fig.~\ref{benchmark}. In this fixed-workflow comparison, {\sc Automat} achieves better performance than all three baseline 
descriptor sets for both target properties. For experimental band-gap prediction, the best baseline is the combined fractional-composition 
and Magpie representation, which gives a test MAE of 0.407~eV. {\sc Automat} reduces this value to 0.352~eV, while also improving the 
test $R^2$ from 0.646 to 0.706. For Curie-temperature prediction, the best baseline is again the combined fractional-composition 
and Magpie representation, with a test MAE of 72.16~K. {\sc Automat} reduces this value to 67.13~K, and increases the test $R^2$ 
from 0.836 to 0.849. These values are in the range of reported state-of-the-art models based on manually curated descriptor sets~\cite{James_tc}.

\subsection{Experimental band-gap prediction}

The \texttt{expt\_gap} dataset contains 4,604 inorganic compositions with experimentally measured band gaps from 
Zhuo \textit{et al.}~\cite{matbench_exp_gap}. The regression task is to predict the experimental band gap from chemical 
composition alone.
Figure~\ref{bandgaprun} shows the descriptor-design trajectory for this task, including the mean cv-MAE across the three 
fixed cross-validation folds and the MAE on the held-out validation set for accepted descriptor updates. The largest 
improvement occurs early in the run, when {\sc Automat} introduces oxidation-state and ionic-balance descriptors. These 
descriptors encode information about possible charge assignments, oxidation-state ranges, charge neutrality, ionic strength, 
and cation--anion partitioning. Their introduction produces a substantial reduction in both the cv-MAE and the held-out 
validation MAE, indicating that the improvement is not restricted to the cross-validation folds used for descriptor acceptance.

Subsequent accepted updates provide smaller gains. The addition of thermodynamic and size-related elemental-property 
statistics, followed by a fractional composition array over 118 elements, further improves the representation, but the magnitude 
of these improvements is smaller than that associated with oxidation-state and charge-balance descriptors. These later additions 
are also less specific to the band-gap problem and resemble more general composition-based feature expansions.

After the ninth accepted update, further reductions in cv-MAE become small and are accompanied by a rapid increase in descriptor 
dimensionality. In particular, later iterations introduce high-dimensional element-co-occurrence features, increasing the descriptor 
size to more than 20,000 components by the end of the run. Although some of these expansions reduce cv-MAE and are therefore 
accepted by the local hill-climbing rule, they do not consistently improve held-out validation performance. Conversely, attempts by 
the agent to remove or compress features worsened cv-MAE and were rejected, while feature expansions were often rewarded by 
the cross-validation criterion. This behavior illustrates a limitation of a purely greedy accept/reject rule: without an explicit complexity 
penalty, the agent may continue to add descriptors even when the resulting representation is no longer preferable for generalization 
or interpretability.
The held-out validation set is therefore essential for selecting a descriptor set from the accepted trajectory. For the band-gap 
task, the descriptor set obtained after iteration 9 provides the best trade-off between validation performance and descriptor 
complexity. This selected representation contains 243 descriptors and is summarized in Table~\ref{tab:gap}. It combines stoichiometric 
statistics, weighted elemental-property statistics, element-family fractions, oxidation-state descriptors, ionic-balance descriptors, 
size and thermodynamic properties, radius-contrast terms, and a fractional composition array.

The descriptor families identified by {\sc Automat} are chemically plausible for band-gap prediction. Band gaps in inorganic compounds 
are strongly influenced by the contrast between electropositive and electronegative elements, the degree of ionic or covalent bonding, 
possible oxidation states, charge transfer, and electronic shell structure~\cite{matbench_exp_gap, Walsh2018}. {\sc Automat} introduced 
descriptors targeting these quantities, and their utility was confirmed through validation feedback. The resulting representation extends 
conventional Magpie-style composition statistics~\cite{magpie, James_tc} with task-specific descriptors that more directly encode bonding, 
ionicity, and charge-balance information.

This run also shows that the generic simplicity criterion used in the original autoresearch implementation is not sufficient 
for descriptor-design problems of this type. In the absence of an explicit descriptor-size constraint, the agent tends to greedily expand 
the feature space. To reduce this behavior, subsequent versions of \texttt{program.md} include guidance on the maximum acceptable 
descriptor dimensionality. The optimal descriptor size is itself problem dependent and could in principle be selected using automated 
feature-selection or feature-importance strategies~\cite{modnet1, modnet2}. In this work, however, we focus on a minimal baseline 
implementation of the autoresearch paradigm and leave systematic descriptor pruning to future work.

\subsection{Curie-temperature prediction}

The \texttt{expt\_Tc} dataset contains 3,638 unique ferromagnetic compounds and their associated Curie temperatures. 
We use the database of Nelson \textit{et al.}~\cite{James_tc}, which aggregates data from the \textit{AtomWork} 
database~\cite{Yamazaki2011}, \textit{Springer Materials}~\cite{Connolly2012}, the \textit{Handbook of Magnetic 
Materials}~\cite{Handbook}, and the book \textit{Magnetism and Magnetic Materials}~\cite{Coey}. This database 
is combined with additional $T_\mathrm{C}$ values manually aggregated by Byland \textit{et al.}~\cite{Valentin-co}, 
which are drawn primarily, although not exclusively, from Co-containing compounds.

Figure~\ref{tcrun} shows the descriptor-design trajectory for the Curie-temperature prediction task. {\sc Automat} 
immediately proposes descriptors based on magnetic chemistry. Despite receiving only a one-line problem description, the agent 
identifies the relevance of transition-metal, rare-earth, actinide, heavy-element, and anion chemistry. Therefore, the first 
accepted descriptor set already contains features designed to emphasize the concentration of magnetic elements and 
chemically relevant element families. Subsequent iterations introduce descriptors that further emphasize the presence of a magnetic sublattice.

The largest improvement in the trajectory occurs when charge-balance features are introduced. A further reduction in the 
held-out validation MAE is observed at iteration 10, coinciding with the introduction of fractional composition descriptors 
over the periodic table. Subsequent accepted updates reduce cv-MAE, but increase the held-out validation MAE, indicating 
overfitting to the training/search set. In contrast to the band-gap run, however, the descriptor dimensionality remains within 
a few hundred features, since this run used an updated \texttt{program.md} file that explicitly instructed the agent to keep 
the total descriptor size below 500. The selected descriptor set corresponds to iteration 10, where it achieves the lowest 
held-out validation MAE of the run.

The selected descriptor set for this task is summarized in Table~\ref{tab:tc}. It contains 261 descriptors, including stoichiometric 
statistics, targeted magnetic-chemistry descriptors, weighted elemental-property summaries, property extrema and spreads, 
magnetic-sublattice descriptors, rare-earth and actinide family fractions, family interaction terms, valence-prior summaries, 
valence-balance terms, periodic-block terms, and a fractional composition array.
Many of these descriptors are directly related to known chemical factors controlling the Curie temperature of permanent 
ferromagnets. The representation emphasizes Fe, Co, magnetic 3d transition metals, rare-earth elements, actinides, selected 
anions, and interactions between magnetic and nonmagnetic sublattices~\cite{Coey}. These terms allow the random forest to 
distinguish compositions dominated by different magnetic sublattices and to encode interactions between transition-metal content, 
rare-earth content, and anion or interstitial chemistry.

The Curie-temperature run also reveals a limitation of the current implementation. Some descriptors duplicate information that 
is already present elsewhere in the feature vector. For example, the fractional concentration of Fe is introduced as part of a 
targeted magnetic-chemistry descriptor family and later appears again in the fractional composition array. Although such 
redundancy is not ideal from an interpretability or compactness perspective, it may also act as a form of implicit feature re-weighting 
by increasing the representation of elements or chemical families that the agent identifies as important.

This behavior likely arises because the agent tends to reason about descriptors in logical blocks. Descriptor updates are therefore often 
implemented as additions, removals, or refinements of entire descriptor families rather than as fine-grained changes to individual 
features. As a result, when the agent attempts to simplify the representation, it may remove a complete descriptor block even if only 
some components of that block are redundant or uninformative. This limits the granularity of the current search procedure and can 
preserve unnecessary feature duplication. More systematic descriptor pruning, redundancy control, or feature-importance-guided 
refinement could therefore improve the compactness and interpretability of future {\sc Automat}-generated representations.

The selected descriptor set improves over all baseline representations. {\sc Automat} achieves a test MAE of 67.13~K, compared 
with 72.16~K for the best baseline, and increases the test $R^2$ from 0.836 to 0.849. These improvements indicate that the 
agent-generated descriptors provide useful task-specific information beyond both the elemental-property statistics in Magpie 
and the explicit elemental identities contained in the fractional composition array.

Overall, these results support the use of autonomous research agents for composition-only descriptor design. Across two distinct 
materials-property prediction tasks, {\sc Automat} identifies chemically meaningful descriptor families and improves random forest 
models beyond established engineered-descriptor baselines. The largest improvements arise when the agent discovers descriptors 
that are closely connected to the target property, such as oxidation-state and charge-balance descriptors for band gaps or 
magnetic-sublattice and family-interaction descriptors for Curie temperatures. In addition to improving predictive performance, the 
generated descriptor trajectories provide an inspectable record of which chemical features are informative for each task, suggesting 
that autoresearch workflows can support both model development and scientific interpretation in materials informatics.

\section{Conclusions}

In this work, we have introduced {\sc Automat}, an autoresearch framework for the autonomous design of compositional 
descriptors for materials-property prediction. The broader goal of automating model construction is well established: 
hyperparameter-optimization tools, AutoML workflows, and auto-sklearn-style pipelines provide powerful methods for 
selecting models, tuning parameters, and constructing ensembles~\cite{automl_book, autosklearn}. However, these 
methods typically operate within a search space defined in advance by the user. {\sc Automat} addresses a complementary 
problem: the automated design of the descriptors themselves. This distinction is important for materials informatics because, 
when a chemical formula is represented only by fractional elemental composition, much of the relevant chemical knowledge 
associated with the constituent atoms must be inferred directly from the data. In low-data experimental settings, this can limit 
predictive performance~\cite{xu2023_small, zhang2018_small}.

{\sc Automat} allows an LLM-based coding agent to propose, implement, test, and refine chemically motivated descriptors 
within a quantitative evaluation loop. In the present study, this capability was deliberately tested in a narrow and controlled 
setting: all models used a random forest architecture, and all descriptors were derived only from chemical composition. Within these constraints, {\sc Automat} generated composition-only descriptors that improved over fractional-composition, Magpie, 
and combined fractional-composition--Magpie baselines for both experimental band-gap predictions and Curie-temperature predictions.

The descriptors generated by {\sc Automat} were not arbitrary numerical transformations of the input formulas. Across both 
tasks, the accepted descriptor families were chemically interpretable and aligned with the target property. For band-gap prediction, 
{\sc Automat} identified oxidation-state and charge-balance descriptors as useful. For Curie-temperature prediction, it generated 
descriptors emphasizing magnetic sublattices, transition-metal content, rare-earth and actinide chemistry, valence balance, and 
periodic-block interactions. These results show that {\sc Automat} can act not only as a performance-optimization tool, but also as 
an inspectable procedure for identifying chemically meaningful features in materials datasets.

The study also reveals important limitations of the current approach. The autoresearch loop used here implements a strict greedy 
accept/reject criterion: descriptor modifications are retained only if they immediately improve the cross-validation optimization metric. 
This makes the descriptor-design trajectory easy to interpret, but it may discard promising intermediate directions that require several 
steps before becoming useful. The results also show that the agent can introduce duplicate or near-duplicate descriptors, which may 
behave partly as implicit feature reweighting. Future versions of {\sc Automat} should therefore incorporate more sophisticated search 
strategies, descriptor de-duplication, pruning, explicit feature weighting, and less greedy acceptance criteria. Existing tools such as 
hyperparameter optimizers, AutoML pipelines, feature-selection methods, and ensemble builders could also be integrated directly 
into the agent loop, allowing the agent to use these methods where appropriate rather than replace them.

A key strength of the autoresearch paradigm is its flexibility and extensibility. {\sc Automat} provides a first baseline framework for 
autonomous descriptor design in materials science, and can serve as a foundation for more advanced agentic workflows. A natural 
next step is to extend {\sc Automat} beyond composition-only descriptors. Many materials properties depend strongly on the crystal 
structure. Incorporating structural descriptors, literature-derived information, or domain-specific simulation outputs would allow the 
same autoresearch paradigm to address a broader class of materials-modeling problems.

More generally, autoresearch is applicable to scientific problems in which progress can be expressed through a clear quantitative 
objective and in which candidate ideas can be implemented, evaluated, and rejected reproducibly. Current LLMs are not specifically 
optimized for autoresearch. Nevertheless, the descriptor-design task studied here requires a combination of scientific knowledge, 
coding ability, and logically grounded iteration. {\sc Automat} therefore provides both a practical tool for automated descriptor design 
and a benchmark for evaluating the ability of future LLM agents to participate in scientific research. Within the setting explored in this 
work, our results demonstrate that an autonomous agent can design task-specific, interpretable compositional descriptors without 
human intervention during the optimization loop.

\section{Acknowledgments}
This work was supported by Enterprise Ireland (contract number CF20242326P).

\section{Code Availability}

The {\sc Automat} code is available at \url{https://github.com/m-cobelli/automat}, together with the descriptors designed in this work.

\bibliographystyle{unsrt}
\bibliography{references}

\end{document}